\documentclass[sigconf,nonacm,9pt]{acmart}

\usepackage{libertine}
\usepackage[utf8]{inputenc}
\usepackage{amsmath}

\usepackage{amssymb,amsfonts}
\usepackage{algorithmic}
\usepackage{listings}
\usepackage{graphicx}
\usepackage{textcomp}
\usepackage{standalone}
\usepackage{xcolor}
\def\BibTeX{{\rm B\kern-.05em{\sc i\kern-.025em b}\kern-.08em
    T\kern-.1667em\lower.7ex\hbox{E}\kern-.125emX}}
\usepackage{booktabs} 
\usepackage{subfigure}
\usepackage{units}
\usepackage{blindtext}
\usepackage{multirow}
\usepackage{xcolor,xspace}
\usepackage{paralist}
\usepackage{mdframed}
\usepackage{pict2e}
\usepackage{mdframed}
\usepackage{color}
\usepackage{threeparttable}
\usepackage{array}
\usepackage{pifont}
\usepackage{url}
\usepackage{comment}
\usepackage{lipsum}
\usepackage{arydshln}
\usepackage{makecell}
\usepackage{flushend}

\usepackage{latexsym}
\usepackage{enumitem}
\usepackage{arabtex}
\usepackage{fancyhdr}
\usepackage[english]{babel}
\usepackage{tabularx}
\input{glyphtounicode}

\usepackage[ruled]{algorithm2e}
\usepackage{algorithmic}

\lstdefinestyle{mystyle}{
    commentstyle=\color{codegreen},
    keywordstyle=\color{blue},
    numberstyle=\small\color{codegray},
    stringstyle=\color{codepurple},
    basicstyle=\fontsize{8}{8}\ttfamily,
    frame=single,
    breakatwhitespace=false,         
    breaklines=true,                 
    captionpos=b,                    
    keepspaces=true,                 
    numbers=none,                    
    numbersep=4pt,                  
    showspaces=false,                
    showstringspaces=false,
    showtabs=false,                  
    tabsize=1,
    xleftmargin=0.03\columnwidth,
    xrightmargin=0.03\columnwidth,
}

\newcounter{protocol}

\usepackage{url}

\usepackage{pgfplots}
\pgfplotsset{compat=1.18}
\usetikzlibrary{patterns}

\usepackage[
    n,
    advantage,
    operators,
    sets,
    adversary,
    landau,
    probability,
    notions,
    logic,
    ff,
    mm,
    primitives,
    events,
    complexity,
    asymptotics,
    keys]{cryptocode}

\usepackage{tikz}          
\usetikzlibrary{arrows, patterns, automata, positioning, shapes}

\newcommand{\ignore}[1]{}

\definecolor{darkgreen}{RGB}{25, 176, 0}

\definecolor{darkorange}{RGB}{207, 100, 0}

\newcommand{\edit}[1]{\textcolor{black}{#1}}

\newcommand{\prv}{{\ensuremath{\sf{\mathcal Prv}}}\xspace}
\newcommand{\vrf}{{\ensuremath{\sf{\mathcal Vrf}}}\xspace}
\newcommand{\RA}{{\ensuremath{\sf{\mathit{RA}}}}\xspace}

\newcommand{\chal}{\textit{Chal}\xspace}

\newcommand{\ACFA}{\textit{{ACFA}}\xspace}

\newcommand{\acron}{\textit{{CARAMEL}}\xspace}
\newcommand{\longacron}{Contention Avoidance in Runtime Auditing with Minimized Execution Latency\xspace}

\newcommand{\cfsize}{\ensuremath{\mathit{CF_{size}}}\xspace}
\newcommand{\ssize}{\ensuremath{\mathit{Slice_{size}}}\xspace}

\renewcommand\adv{\ensuremath{\sf{\mathcal Adv}}\xspace}

\newcommand{\metadata}{\textit{METADATA}\xspace}
\newcommand{\er}{\textit{AER}\xspace}

\newcommand{\cfaud}{\textit{CF-Aud}\xspace}
\newcommand{\cfatt}{\textit{CF-Att}\xspace}

\newcommand{\topslice}{\mathit{Slice_{top}}\xspace}
\newcommand{\bottomslice}{\mathit{Slice_{bot}}\xspace}

\newcommand{\logptr}{\ensuremath{Log_{ptr}}\xspace}
\newcommand{\vrfokay}{\mathit{Vrf_{acc}}\xspace}

\newcommand{\logfull}{\ensuremath{log_{full}}\xspace}
\newcommand{\logstate}{\textit{$Trigger_{state}$}\xspace}

\newcommand{\cflog}{\ensuremath{CF_{Log}}\xspace}
\newcommand{\cflogs}{\ensuremath{CF_{Logs}}\xspace}
\newcommand{\cfslice}{\ensuremath{CF_{Slice}}\xspace}
\newcommand{\cfslices}{\ensuremath{CF_{Slices}}\xspace}

\newlist{myitemize}{itemize}{1}
\setlist[myitemize]{
    label=$\bullet$,
    leftmargin=*,
    nosep,
}

\newlist{myenumerate}{enumerate}{1}
\setlist[myenumerate]{
  label=(\arabic*),
  leftmargin=*,
  nosep,
}

\lstdefinestyle{CStyle}{
    backgroundcolor=\color{white},   
    commentstyle=\color{mGreen},
    keywordstyle=\color{blue},
    numberstyle=\small\color{black},
    stringstyle=\color{blue},
    basicstyle=\ttfamily\footnotesize,
    breakatwhitespace=false,         
    breaklines=true,                 
    captionpos=b,                    
    keepspaces=true,                 
    numbers=left,                    
    numbersep=5pt,      
    frame=single,
    xleftmargin=1.75em,
    xrightmargin=1.75em,
    framexleftmargin=1em,
    captionpos=b,
    showspaces=false,                
    showstringspaces=false,
    showtabs=false,                  
    tabsize=2,
    float=tp,
    language=C
}

\begin{document}

\title{Boosting Device Utilization in Control Flow Auditing}

\author{Alexandra Lengert}
\affiliation{%
  \institution{University of Zurich}
  \city{Zurich}
  \country{Switzerland}
}

\author{Adam Ilyas Caulfield}
\affiliation{%
  \institution{University of Waterloo}
  \city{Waterloo, Ontario}
  \country{Canada}
}

\author{Ivan De Oliveira Nunes}
\affiliation{%
  \institution{University of Zurich}
  \city{Zurich}
  \country{Switzerland}
}

\renewcommand{\shortauthors}{Lengert et al.}

\begin{abstract}

Micro-Controller Units (MCUs) are widely used in safety-critical systems, making them attractive targets for attacks. This calls for lightweight defenses that remain effective despite software compromise. Control Flow Auditing (\cfaud) is one such mechanism wherein a remote verifier (\vrf) is guaranteed to received evidence about the control flow path taken on a prover (\prv) MCU, even when \prv software is compromised.
Despite promising benefits, current \cfaud architectures unfortunately require a ``busy-wait'' phase where a hardware-anchored root-of-trust (RoT) in \prv retains execution control to ensure delivery of control flow evidence to \vrf. This drastically reduces the CPU utilization on \prv.

In this work, we addresses this limitation with an architecture for \longacron (\acron). \acron is a hardware-software RoT co-design that enables \prv applications to resume while control flow evidence is transmitted to \vrf. This significantly reduces contention due to transmission delays and improves CPU utilization without giving up on security. Key to \acron is our design of a new RoT with a self-contained (and minimal) dedicated communication interface. \acron's implementation and accompanying evaluation are made open-source. Our results show substantially improved CPU utilization at a modest hardware cost.

\end{abstract}

\maketitle

\section{Introduction}\label{sec:intro}

The use of embedded devices (a.k.a., IoT or ``smart'' devices) is rapidly increasing. They are crucial parts of modern systems with various tasks, including safety-critical operations~\cite{shekari2021mamiot,alrawi2019sok,soltan2018blackiot}. Embedded devices are commonly implemented using Micro-Controller Units (MCUs) designed for energy efficiency and low cost, making them useful for lengthy remote deployments. However, these resource constraints lead to limited security features.

In this landscape, Remote Attestation (\RA)~\cite{nunes2024toward} 
offers an inexpensive security mechanism to remotely assess software integrity. \RA is a challenge-response protocol in which a Verifier (\vrf) challenges a remotely deployed Prover MCU (\prv) to provide cryptographic proof of their currently installed code. By inspecting the proof, \vrf can remotely decide on \prv's code integrity.
Yet, in standard \RA, \vrf remains oblivious to attacks that do not modify code~\cite{cflat}.
For example, an adversary (\adv) can launch control flow hijacking or code-reuse attacks~\cite{jop,rop},
corrupting branch instructions (e.g., calls, jumps, returns) to execute unintended functions, skip security-critical code, or other arbitrary (malicious) behavior without detection.
%
Control Flow Integrity (CFI) mechanisms~\cite{cfi-survey-1} offer a local countermeasure, but often they neither verify complete control flow paths nor produce authenticated path evidence, preventing \vrf from subsequent examination to determine attack root causes.

Control Flow Attestation (\cfatt)~\cite{sok_cfa_cfi} extends standard \RA to provide \vrf with authenticated control flow path evidence.
In \cfatt, a runtime trace of the control flow path (\cflog) taken during the attested program's execution is also included in the cryptographic proof sent to \vrf. 
Upon receiving \cflog, \vrf can inspect it to determine whether the program was executed correctly or if remedial actions are required.

Most previous \cfatt are best-effort, assuming that \cflog would reach \vrf, as a persistent lack of a response to \vrf requests for evidence suggests an issue (e.g., a compromised state) with \prv. 
However, \adv may deliberately not send \cflog to conceal the control flow path that led to the compromise, revealing an anomalous state, but preventing analysis of \cflog for vulnerability (root cause) identification.
To address this, Control Flow Auditing (\cfaud)~\cite{acfa,traces} was proposed to ensure reliable delivery of \cflog. Additionally, a \cfaud RoT can guarantee to \vrf the ability to take corrective action when an invalid execution path is detected.

\cfaud obtains its guarantees by actively triggering a (hardware-supported) RoT to generate and transmit runtime evidence (containing \cflog) to \vrf upon key events at runtime (e.g., when \prv storage for \cflog is full, requiring transmission before adding new entries).
Current \cfaud RoTs ``busy-wait'' in a secure state after evidence transmission until receiving an acknowledgment from \vrf. This ensures that a compromised \prv cannot tamper, delete, or replace \cflog before it has reached \vrf. Clearly, this approach diminishes \prv utilization. Indeed, prior work~\cite{acfa} acknowledged trade-offs between ``best-effort'' \cfatt vs. allowing untrusted execution to resume after a maximum waiting period in \cfaud. Yet, it is thus far unclear how \cflog transmission and untrusted code resumption can be achieved simultaneously.

This work tackles the aforementioned challenge with \acron: an architecture for \longacron.
\acron's fundamental shift from prior RoT models lies in a dedicated and RoT-integrated trusted communication interface. Notably, this integration requires addressing non-trivial challenges. In particular, \cflog's joint memory+transmission management requires the creation of partial authenticated report units (denoted \cfslice). This ensures that new entries, added to \cflog by the resumed application, can still be securely stored as a new \cfslice while prior reports/responses to/from \vrf are transmitted, thus avoiding contention due to lack of storage for \cflog.
In sum, our anticipated contributions are:
\begin{myitemize}
    \item \acron Design: we design a hardware-software RoT architecture for \cfaud  that maintains security and boosts \prv utilization during \cfaud cycles. \acron low-cost hardware records attested program control flow paths, triggers the generation and parallel transmission of reports, 
    and handles reliable communication with \vrf. \acron's software is responsible for cryptographic operations, which are relatively expensive to implement in hardware.
    \item \acron Open Implementation and Evaluation: we implement and analyze cost and security of \acron. As a case study, we implement a working prototype of \acron deployed on the Basys3 FPGA prototyping board (available at~\cite{caramelrepo}) and evaluate end-to-end runtime auditing instances of real-world sensor applications ~\cite{ultsensor,tempsensor,opensyringe,rover}. Our results confirm \acron increases CPU utilization with low hardware cost.
\end{myitemize}

\section{Background: Attestation \& Auditing}\label{sec:cfaud}


Runtime Attestation methods, including \cfatt~\cite{cflat,iscflat,tinycfa,scarr, recfa,oat,blast,traces,acfa,lofat,litehax,atrium,enola}, are concerned with remotely verifying runtime properties beyond the presence of the correct code in \prv's program memory.
\cfatt typically involves the following steps:
\begin{myenumerate}
    \item \vrf requests \cfatt from \prv by sending an attestation challenge (\chal);
    \item After receiving \chal, \prv executes its software while an RoT on \prv records an execution trace ($T$);
    \item \prv's RoT computes an authenticated measurement ($\sigma$) over \prv's program memory, $T$, and \chal to produce report $H = [\sigma, T]$, which it then sends to \vrf;
    \item Upon receiving $H$, \vrf validates it by comparing $\sigma$ to its expected value and examining $T$ to determine if $T$ represents expected program execution properties.
\end{myenumerate}

In \cfatt and \cfaud, $T$ corresponds to \cflog.
The authenticated measurement ($\sigma$) computed in Step 3 is typically implemented as a Message Authentication Code (MAC) or digital signature. 
The secret key used for the measurement is kept inaccessible to all software on \prv, except for the isolated RoT implementation, typically requiring hardware support. Current architectures build \cflog by either (1) instrumenting the program binary with calls to a logging software inside a Trusted Execution Environment (TEE) or (2) custom hardware modules that detect and log control flow transfers.


Early \cfatt techniques~\cite{cflat,lofat,atrium} send \vrf a single hash digest produced by computing a hash-chain of control flow transfers, yielding a fixed-sized representation of the program path.
Since deriving \cflog from the hash becomes infeasible as the number of control flow transfers in a program grows~\cite{aliasing}, most of the recent \cfatt techniques transmit \cflog \textit{verbatim} to \vrf~\cite{acfa,traces,tinycfa,litehax,oat,recfa,blast,ari,scarr}.
Transmission and storage costs of \cflogs, can be reduced for by compressing data written to \cflog without loss of information~\cite{speccfa,oat,litehax,acfa,traces,recfa,blast,ari} or by transmitting partial \cflog slices when dedicated \cflog storage has reached capacity~\cite{litehax,scarr,acfa,traces}.

\cfatt mechanisms were previously limited by their \textit{passive} nature, relying on a compromised \prv to participate in a \cfatt protocol and send the runtime evidence. However, there is no guarantee that \adv who has full control over \prv software will allow \cflog to be sent.
The latter prevents \vrf from (1) receiving evidence of the exact path deviation, (2) learn any resulting consequences, or (3) remediating attack root causes.

\cfaud architectures~\cite{acfa,traces} incorporate mechanisms to actively interrupt the attested application's execution to transmit \cflog reliably upon significant runtime events (e.g., storage full, timeout, execution completed).
While waiting for \vrf confirmation of evidence receipt, current \cfaud RoTs retain execution control of \prv, ensuring that application execution only resumes once \vrf confirms its trust in the evidence sent by \prv. However, as noted in Sec.~\ref{sec:intro}, this results in loss of \prv utilization, especially for longer programs that require several partial \cflog transmissions. 

\section{\acron Overview}\label{sec:overview}

\begin{figure}[t] 
    \centering
    \includegraphics[width=.8\columnwidth]{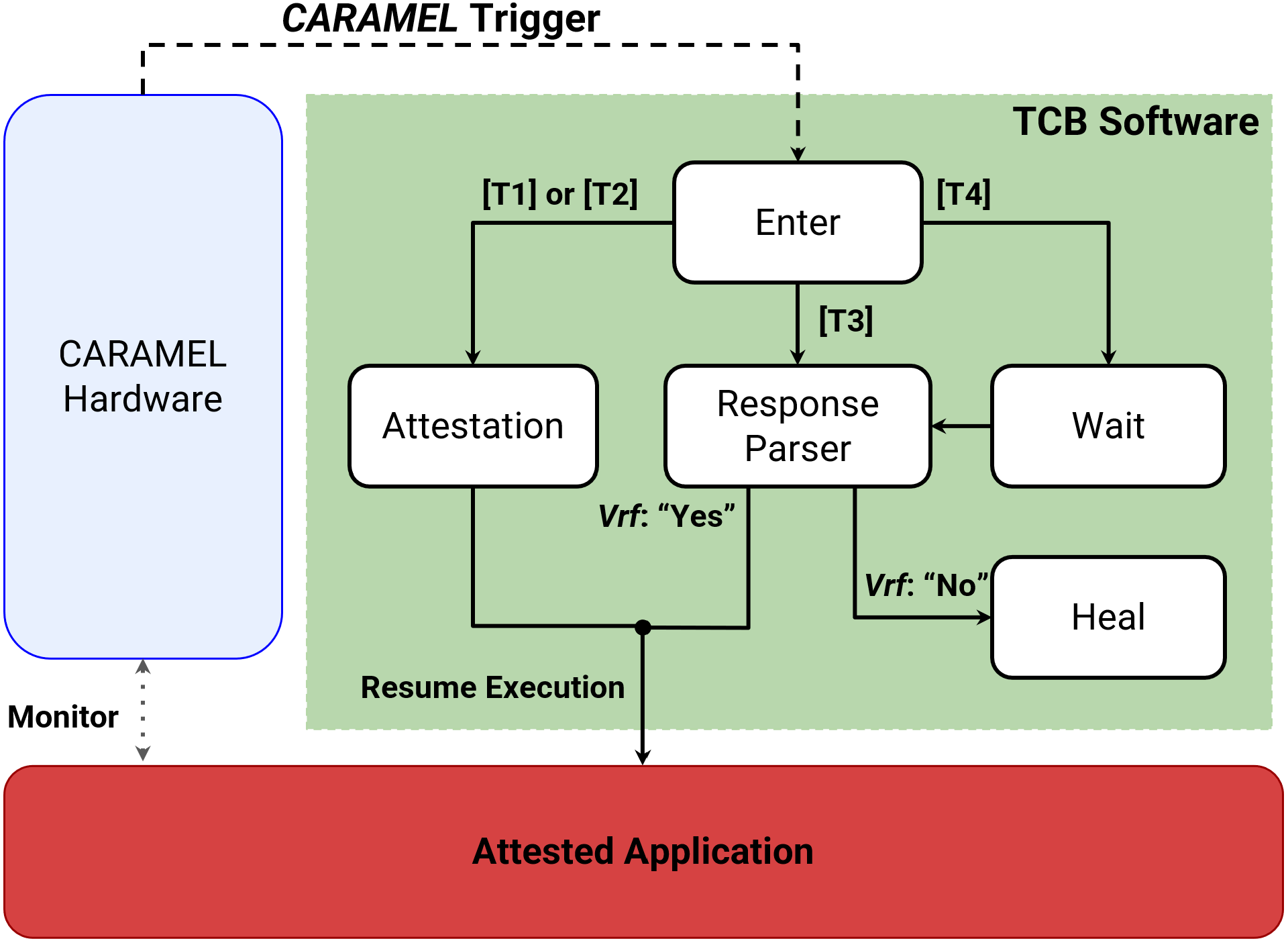}
    \vspace{-1em}
    \caption{\acron high-level workflow} 
    \label{fig:system}
    \vspace{-2em}
\end{figure}

\acron avoids \cflog contention in \cfaud by partitioning \cflog storage into two adjacent \cfslices. During runtime, \acron hardware detects when the first \cfslice is ready for transmission and initiates transmission via an RoT-integrated communication interface. 
\acron allows the application's execution to resume in parallel (which in turn starts to fill up a second \cfslice) while waiting for a response from \vrf. It generates a secure (i.e., non-maskable) interrupt when a message is received through the RoT communication interface to begin its processing.
\acron software Trusted Computing Base (TCB) is only required to (1) compute/validate authentication tokens and (2) execute \vrf-specified remediation action(s), in case of compromise detection. The complete workflow is shown in Fig.~\ref{fig:system}.

During execution, \acron actively triggers its TCB via a custom non-maskable interrupt upon the following events:

\noindent \textbf{[T1]} the audited application concludes its execution;
    
\noindent \textbf{[T2]} a \cfslice is full and ready for transmission to \vrf;
    
\noindent \textbf{[T3]} a message is received by the RoT communication interface; 
    
\noindent \textbf{[T4]} \cflog storage (both \cfslices) fills up while no \vrf  response is received. In this case, \acron must wait before \edit{reusing \cflog.}

Per Fig.~\ref{fig:system}, \acron TCB's execution path depends on the trigger source. 
When triggered by \textbf{[T1]} or \textbf{[T2]}, an attestation report must be created.
After the report is ready, the TCB exits to resume the application execution.
\acron hardware ensures new control flow transfers do not overwrite any \cfslice that is pending verification by \vrf 
(further details discussed in Sec.~\ref{sec:components}).

When triggered by \textbf{[T3]}, \acron has received a message through its communication interface, and thus it authenticates and parses the message. 
If \vrf approved the previous report, the TCB exits and resumes the application, making the previous \cfslice memory available for reuse. Alternatively, a \vrf-configurable remediation is triggered if the report indicates a compromise.

If TCB is triggered by \textbf{[T4]}, \acron has not received a timely response from \vrf, both \cfslices have been used, and it must enter a secure ``busy-wait'' state as a last resort, falling back to the same mechanism used by prior \cfaud architectures by default~\cite{acfa,traces}. The TCB stays in this state until a message is received from \vrf approving the pending \cfslice.



\section{\acron Design Details}\label{sec:design}




\subsection{System \& Adversary Models}\label{subsec:adv}


We focus on resource-constrained MCUs (e.g., AVR ATmega~\cite{atmega} and TI MSP430~\cite{tiMSP430} families). MCUs within this class are typically equipped with 8- or 16-bit cores that run on low clock frequencies (1-16 MHz). Due to a lack of MMUs or virtualization support, they run software at ``bare metal'' (i.e., without virtual-to-physical address translations). \acron prototype is built on the openMSP430 MCU implemented in Verilog~\cite{openmsp430}. We chose openMSP430 due to its public availability, though we believe \acron design to be equally applicable to other MCUs.

We consider that \adv can fully compromise any software on \prv that is not explicitly protected by trusted hardware. \adv can exploit vulnerabilities in \prv software to inject code~\cite{francillon2008code}, hijack the application's control flow~\cite{ma2023ret2ns}, or perform code-reuse attacks~\cite{rop,jop}. Additionally, \adv can manipulate interrupts and their associated routines, if unprotected. \adv may also discard, inject, or attempt to modify messages in the network between \prv and \vrf. In line with prior works~\cite{cflat,oat,litehax,acfa,pfb,casu,delegated}, physical attacks are out of scope as they require orthogonal measures~\cite{obermaier2018past}.

\subsection{\acron Architecture}\label{sec:arch}

\begin{figure}[t] 
    \centering
    \includegraphics[width=\linewidth]{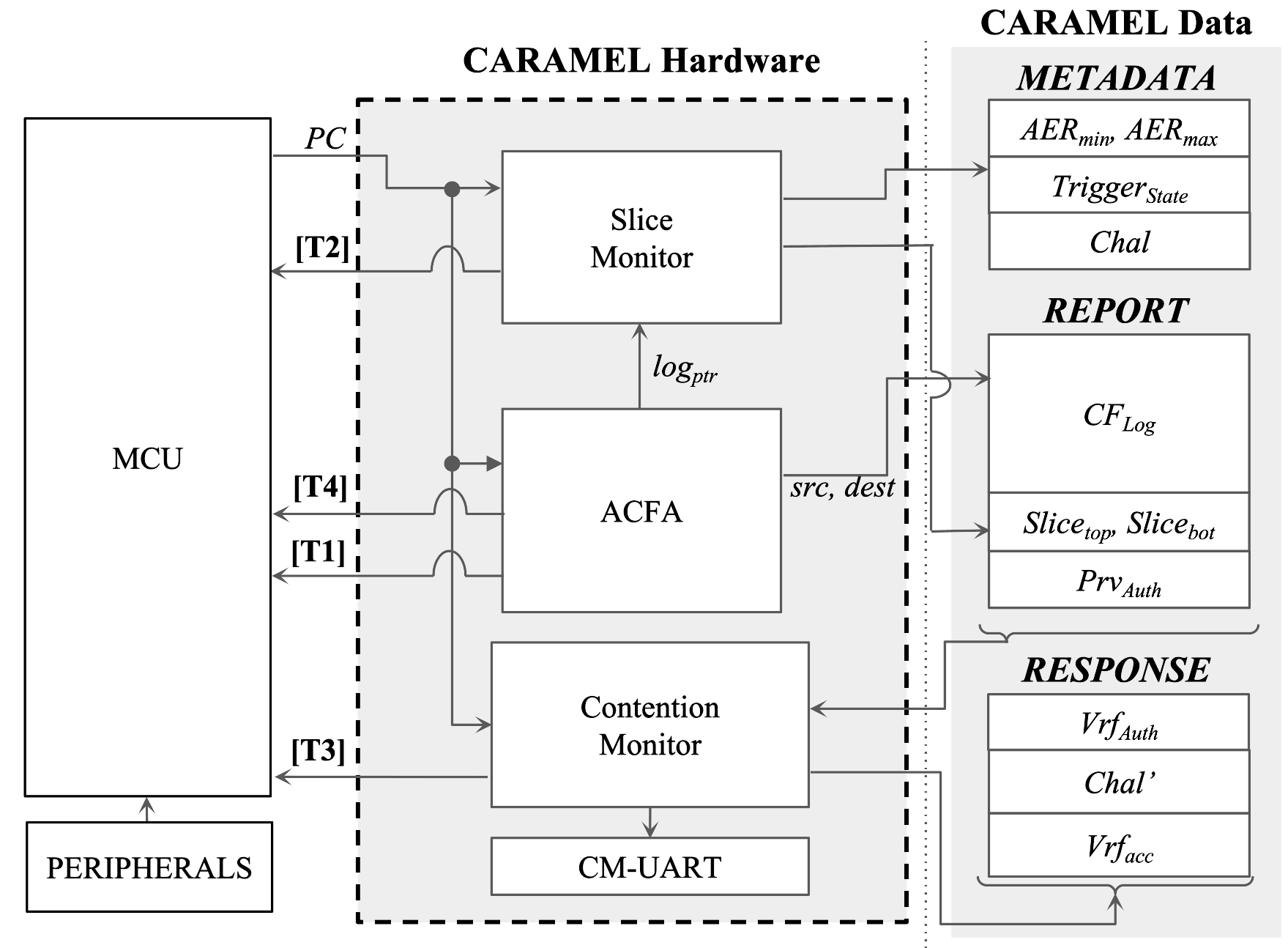}
    \vspace{-2.5em}
    \caption{\acron system architecture } 
    \vspace{-1.5em}
    \label{fig:arch}
\end{figure}

\acron interfaces with the MCU core and monitors some of its signals, as depicted in Fig.~\ref{fig:arch}.
MCU memory consists of program (PMEM) and data memory (DMEM). As shown in Fig.~\ref{fig:system}, PMEM is divided between \acron's hardware-protected TCB code and untrusted software (\textit{S}) containing applications, including an application to be audited. The PMEM section used to store the audited application is called Audited Executable Region (\er). \er boundaries can be configured by \vrf if specified as part of the initial \cfaud request. 
\acron data in DMEM is used to exchange data between \acron's hardware and TCB (more details in Sec.~\ref{sec:components}). The rest of DMEM is available for \textit{S} allocation.

\acron hardware consists of three main modules: the underlying auditing architecture \ACFA~\cite{acfa}, the Slice Monitor, and the Contention Monitor.
\acron modifies \ACFA for compatibility with other \acron components (described further in Sec.~\ref{sec:components}).
\acron uses \ACFA sub-modules to:
\begin{myitemize}
    \item generate \textbf{[T1]} and \textbf{[T4]} to trigger execution of \acron's TCB;
    \item protect \acron's TCB code and its execution from direct tampering by \textit{S};
    \item detect control flow transfers during \er's execution and log them into \cflog. 
\end{myitemize}
%

\acron's Slice Monitor monitors \cflog and sets \textbf{[T2]} trigger when a \cfslice is full (i.e., a partial report should be generated for \vrf).
The Contention Monitor checks for when a partial \cfaud report is ready to be transmitted via its dedicated communication interface: the Contention Monitor UART (CM-UART). The Contention Monitor also receives responses from \vrf and sets \textbf{[T3]} to trigger response parsing (described further in Sec. \ref{sec:components}.)

 


\subsection{TCB Software}\label{sec:tcb}

\acron's TCB implements five procedures, as shown in Fig.~\ref{fig:system}.
The \texttt{Enter} procedure determines the execution path based on the trigger source.
The \texttt{Heal} procedure is configurable by \vrf based on the appropriate remediation for the deployment context (some examples include shut down, reset, erase DMEM, or software update). The \texttt{Wait} procedure is implemented as a ``busy-wait'' loop.  
The \texttt{Attestation} and \texttt{Response} \texttt{Parser} procedures incorporate components from VRASED~\cite{vrased} formally verified RA architecture. VRASED routines are used to produce authenticated memory measurements and to authenticate \vrf messages. This is because ACFA itself is built upon VRASED. 

The TCB exit context affects \acron hardware behavior:
\begin{myenumerate}
    \item if exiting after \texttt{Response} \texttt{Parser} procedure, an authenticated \vrf message was received and \vrf approved the previous report (thus, \acron hardware can now overwrite/reuse the memory corresponding to the accepted \cfslice);
    \item if exiting after the \texttt{Attestation} procedure, an authentication token for the current report has been fully produced. Thus, hardware should start transmitting an authenticated report.
\end{myenumerate}
%
%
For \acron hardware to be aware of these conditions, the TCB redirects its execution to two corresponding trampoline instructions at the following fixed addresses:
\begin{myenumerate}
    \item \textit{$accepted_{addr}$}: visited when \vrf was authenticated and \vrf accepted the previous report;
    \item \textit{$send_{addr}$}: visited when the authentication token has been generated and is ready for transmission.
\end{myenumerate}

\acron hardware monitors the program counter ($PC$). PC contains the address of the currently executing instruction.
Based on $PC$, the Slice Monitor detects visits to \textit{$accepted_{addr}$} to release the prior \cfslice memory for reuse.
The Contention Monitor begins transmitting data for the report while the authentication token is computed, and waits for visits to \textit{$send_{addr}$} to continue transmission of the corresponding token.
Sec.~\ref{sec:components} discusses further details on the exit conditions checked by \acron hardware.


\subsection{\acron Data}\label{sec:data}

The dedicated memory for \acron data contains three subdivisions: \metadata, \textit{REPORT}, and \textit{RESPONSE}, as depicted in Fig.~\ref{fig:arch}.


The \textbf{\metadata} region contains data that is used internally by \acron. It has the following components:

\begin{myitemize}
    \item $\er_{min}$, $\er_{max}$: store the bounds of \er, specified by its minimum and maximum memory address, respectively;
    \item \logstate: stores the trigger source that caused the latest TCB to execution; 
    \item \chal: stores the latest challenge received from \vrf.
\end{myitemize}

The \textbf{\textit{REPORT}} region contains all data to be eventually sent to \vrf alongside the metadata in audited execution report(s):
\begin{myitemize}
    \item \cflog: the memory region for storing control flow transfers;
    \item $\topslice, \bottomslice$: the bounds of the \cfslices in \cflog that should be used for the auditing report, denoting the top (first address) and bottom (last address) of a \cfslice, respectively;
    \item $\prv_{Auth}$: memory for storing the report authentication token computed by the TCB's \texttt{Attestation} procedure.
\end{myitemize}

The \textbf{\textit{RESPONSE}} region contains the data that should be received as a part of \vrf's response. Within it are the following:
\begin{myitemize}
    \item $\vrf_{Auth}$: token sent by \vrf to authenticate the message;
    \item $\chal'$: fresh attestation challenge for the next report;
    \item $\vrf_{check}$: byte denoting whether \vrf accepted ($\vrf_{check}=1$) or rejected ($\vrf_{check}=0$) the report. 
\end{myitemize}

\subsection{Specification of \acron Components}\label{sec:components}

This subsection defines the hardware specifications for \acron's internal sub-modules depicted in Fig.~\ref{fig:arch}. Several variables that reference \cflog bounds are incremented to wrap around rather than overflowing the size of \cflog. We describe this by defining the incrementing function $\mathtt{incr(\cdot)}$ in Fig.~\ref{fig:logstate}.

\subsubsection{\textbf{\ACFA Module}}

\ACFA's hardware~\cite{acfa} interfaces directly with \acron's Slice Monitor, writes to \textit{REPORT}, and outputs 
\textbf{[T1]} and \textbf{[T4]}.
\edit{\textbf{[T1]} is set when $PC$ reaches the last instruction of \er ($\er_{max}$)}
\textbf{[T4]} originates from \ACFA's signal \logfull. \ACFA maintains a signal (\logptr) that tracks the total data added to \cflog thus far.
In original \ACFA, \logfull was set when the next \logptr value (\logptr+2) equaled  \cflog's max size (\cfsize). In \acron, \logptr is defined to wrap when reaching \cfsize, and \cflog is considered full when no \cfslice is available. Hence, we modify this definition to set \logfull \edit{when \logptr+2 equals $\topslice$, as shown in Fig.~\ref{fig:logstate}.}

\begin{figure}[t]
\scriptsize
\fbox{
    \parbox{0.95\columnwidth}{     
        \textbf{\underline{Definition:}} Wrapping increment 
        {
            \begin{equation*}
                \mathtt{incr}(base, \delta):= \mathit{base+\delta \mod \cfsize}
            \end{equation*}
        }

        \textbf{\underline{Definition:}} Modified definition of $\logfull$:
        \begin{equation*}
                \mathit{\logfull := (\mathtt{incr}(\logptr,2) = \topslice)}
            \end{equation*}
    
        \textbf{\underline{HW Specification:}} TCB Trigger \textbf{[T1]} and \textbf{[T4]}
        
        {
            \begin{equation*}
                \mathit{(PC = \er_{max}) \rightarrow [T1]}
            \end{equation*}
            \begin{equation*}
                \mathit{\logfull \rightarrow [T4]}
            \end{equation*}
        }
    }
}
\vspace{-1.5em}
\caption{Increment Definition and Modification to \ACFA}
\vspace{-2em}
\label{fig:logstate}
\end{figure}

\subsubsection{\textbf{Slice Monitor}}

\edit{The Slice Monitor has two main roles:}
\begin{myenumerate}
    \item sets \textbf{[T2]} when TCB should generate a partial report;
    \item tracks which \cfslice is used for the attestation report by updating $\topslice$ and $\bottomslice$. 
\end{myenumerate}

\begin{figure}[t]
\scriptsize
\fbox{
    \parbox{0.95\columnwidth}{
        \textbf{\underline{HW Specification:}} Maintain trigger enumeration (\logstate) 
        {
            \begin{equation*}
                \logstate := select\{[T1], [T2], [T3], [T4]\}
            \end{equation*}
        }

        \textbf{\underline{HW Specification:}} The current \cfslice is full
        {
            \begin{equation*}
                \mathit{(\logptr = bound_{low}) \rightarrow slice_{full}}
            \end{equation*}

        }
    
        \textbf{\underline{HW Specification:}} Track if \vrf accepted the previous report (\textit{$\vrfokay$}) 
        {
        \begin{equation*}
            \mathit{\vrfokay} := 
            \left\{
            \begin{array}{ll}
                 \mathit{1}  & \text{if } \mathit{PC = accepted_{addr}} \\
                 \mathit{0} & \text{if }  [T2] \\
                 \mathit{\vrfokay} & \text{otherwise}
            \end{array}
            \right.
        \end{equation*}
        }

        \textbf{\underline{HW Specification:}} TCB Trigger [T2] 
        {
            \begin{equation*}
                \mathit{slice_{full} \land \vrfokay \rightarrow [T2]}
            \end{equation*}
        }

        \textbf{\underline{HW Specification:}} Update 
        pointer to \cfslice top ($\topslice$)
        {
            \begin{align*}
                \mathit{(PC = accepted_{addr})} &\mathit{\rightarrow (\topslice := \mathtt{incr}(\topslice,\ssize))} 
            \end{align*}
        }
        \textbf{\underline{HW Specification:}} Update internal monitor of \cfslice limit ($bound_{low}$)
        {
            \begin{align*}
                slice_{full} \xspace \land \neg \log_{full} \rightarrow (bound_{low} := \mathtt{incr}(bound_{low}, \ssize))
            \end{align*}
        }
        \textbf{\underline{HW Specification:}} Update 
        pointer to \cfslice bottom ($\topslice$)
        {
            \begin{align*}
                \mathit{[T1] \land \vrfokay\xspace} &\mathit{\rightarrow (\bottomslice := \logptr)}\\
                [T2] &\mathit{\rightarrow (\bottomslice := bound_{low})}
            \end{align*}
        }
    }
}
\vspace{-1.5em}
\caption{Slice Monitor hardware specification}
 \vspace{-2em}
\label{fig:slicefull}
\end{figure}

\edit{The Slice Monitor (see hardware specification in Fig.~\ref{fig:slicefull}) maintains bit vector $\logstate$ to keep track of the current active trigger, as mentioned in Sec. \ref{sec:data}. }
%

The remaining specification relates to monitoring the \cfslice that is in use for logging and transmission, \edit{via three pointers:}
\begin{myenumerate}
    \item $\topslice$, pointing to the top limit of the \cfslice that should be used in the next report;
    \item $\bottomslice$, pointing to the bottom of the \cfslice that should be used in the next report.
    \item $bound_{low}$, pointing to the bottom limit of the \cfslice that is currently being written to by \acron hardware;
\end{myenumerate}
The pointers ($\topslice$,$\bottomslice$) are in memory and read by TCB software (as shown in Fig.~\ref{fig:arch}), whereas $bound_{low}$ is used internally.
To track when the limit of the current \cfslice has been reached, Slice Monitor sets \textit{$slice_{full}$} when \logptr reaches $bound_{low}$.

$\vrfokay$ represents whether the previous report is pending ($\vrfokay=0$) or accepted ($\vrfokay=1$). 
The Slice Monitor clears $\vrfokay$ when \textbf{[T2]} has been set (i.e., a new report was generated).
It sets $\vrfokay$ when $PC$ =  \textit{$accepted_{addr}$}, signaling that \acron's TCB has received and authenticated \vrf's response (recall Sec.~\ref{sec:tcb}). 
When both \textit{$slice_{full}$} and $\vrfokay$ are set, the Slice Monitor sets \textbf{[T2]}.

The Slice Monitor updates the \cfslice boundary points based on the previously described signals.
$\topslice$ alternates between the tops of the two \cflog halves (i.e., either points to index 0 or \ssize) when \vrf has accepted the report (i.e., $PC$ = \textit{$accepted_{addr}$}).

The Slice Monitor updates $bound_{low}$ to point to the bottom of the next slice when $slice_{full}$ and $\neg log_{full}$ (i.e., when the current \cfslice is full, but the next one is still available for use). 

$\bottomslice$ is updated upon the \textbf{[T1]} and \textbf{[T2]} triggers to ensure the next report references the latest \cfslice. 
When \textbf{[T1]} has occurred and $\vrf_{acc}$ is set, $\bottomslice$ is set to \logptr, to account for \textbf{[T1]} occurring in the middle of \cfslice boundaries. 
When \textbf{[T2]} has occurred, $\bottomslice$ is set to $bound_{low}$. 
\subsubsection{\textbf{Contention Monitor}}

\edit{The Contention Monitor 
is responsible for controlling all communication with \vrf. It reads from \textit{REPORT} and writes to \textit{RESPONSE}, which triggers \textbf{[T3]}. 
It transmits the report in parallel to software execution and is configured to send new bytes upon a pulse based on the CM-UART's baud rate ($trigger_{tx}$). }

%
Fig.~\ref{fig:cont_mon_tx} shows the specification for the Contention Monitor to track which triggers require a transmission, to iterate through \textit{REPORT} memory for transmission, and to pass bytes from \textit{REPORT} to the CM-UART transmit buffer. It is synchronized on $trigger_{tx}$ and implemented with the following internal signals and registers:
\begin{myitemize}
    \item (\textit{$t1_{pend}, t2_{pend}$}): registers corresponding to whether a transmission is required due to trigger \textbf{[T1]} or \textbf{[T2]}, respectively;
    \item \textit{$read_{idx}$}: an index to read from \textit{REPORT};
    \item \textit{$start_{\cflog}$}: flag to start sending data from \cflog;
    \item \textit{$finsh_{\cflog}$}: flag denoting finished \cflog data transmission;
    \item \textit{$start_{rem}$}: flag to start sending the remaining \textit{REPORT} data;
    \item \textit{$data_{TX}$}: the transmission buffer of CM-UART. 
\end{myitemize}

The Contention Monitor begins transmitting after \textit{$tx_{pend}$}, which is the accumulation of \textit{$t1_{pend}$} and \textit{$t2_{pend}$}. These registers are not cleared until their corresponding reports have been sent. Since \textbf{[T1]} may occur during transmission of \textbf{[T2]} report, clearing of \textit{$t1_{pend}$} is also conditioned on \textit{$t2_{pend}$} already being cleared.


The relevant \cfslice(s) are transmitted during computation of $\prv_{auth}$ to improve performance. The Contention Monitor initializes this by setting $start_{\cflog}$ when:
\begin{myitemize}
    \item a transmission has been registered ($tx_{pend}$ is set);
    \item \vrf has accepted the previous report ($\vrf_{acc}$ is set);
    \item and there is no ongoing transmission ($start_{\cflog}$ is cleared);
\end{myitemize}
Additionally at this moment, the Contention Monitor sets \textit{$read_{idx}$} to start reading from $\topslice$.
The Contention Monitor detects the completed transmission of \cfslice(s) when $read_{idx}$ equals $\bottomslice$, at which point \textit{$start_{\cflog}$} is cleared and \textit{$finish_{\cflog}$} is set.

The remaining \textit{REPORT} data is transmitted when \textit{$finish_{\cflog}$} is set and the \textit{$\prv_{auth}$} has been generated.
The latter is detected when $PC$ equals \textit{$send_{addr}$} (recall Sec.~\ref{sec:tcb}). When these two conditions occur, the Contention Monitor sets $start_{rem}$ to initiate transmission of remaining \textit{REPORT} data, and it sets \textit{$read_{idx}$} to the starting point of \textit{$\prv_{auth}$} in \textit{REPORT}: the base of the \cflog (depicted in Fig.~\ref{fig:arch}).

To complete the transmission of a byte, the Contention Monitor indexes \textit{REPORT} memory using \textit{$read_{idx}$} to obtain the next byte, and is passes it to the CM-UART data transmission buffer ($data_{TX}$).




The specification for the Contention Monitor to handle \vrf's response is outlined in Fig.~\ref{fig:cont_mon_rx}. The CM-UART passes the following receive signals to the Contention Monitor:
\begin{myenumerate}
    \item \textit{$trigger_{rx}$}: set when the CM-UART has finished receiving a byte;
    \item \textit{$data_{RX}$}: the received byte value itself.
\end{myenumerate}
The Contention Monitor maintains \textit{$write_{idx}$} as an internal write index for writing bytes to \textit{RESPONSE} memory region.
With each received byte, \textit{RESPONSE} region is updated and \textit{$write_{idx}$} is incremented. 
When \textit{$write_{idx}$} equals the \textit{RESPONSE} size limit (\textit{$RESP_{Size}$}), receiving is concluded, \textbf{[T3]} is set, and \textit{$write_{idx}$} is cleared.



\section{Prototype \& Evaluation}



\edit{We synthesize \acron hardware using the Xilinx Vivado toolset, implementing all \acron sub-modules in Verilog according to the logic in Sec.~\ref{sec:design}. Our open-source prototype~\cite{caramelrepo} is based on openMSP430 core~\cite{openmsp430}. We configure openMSP430 with 24 KB of PMEM and 30 KB of DMEM, retain its baseline GPIO, Timer, and UART peripherals, and add a second UART instance as the CM-UART operating at 115200 baud as part of \acron’s RoT (Sec.~\ref{sec:components}). Within \acron’s TCB, \vrf authentication and authenticated integrity measurement (see Sec.~\ref{sec:cfaud}) employ a SHA-256–based HMAC from the formally verified HACL$^{*}$ library~\cite{hacl}.}

    

\begin{figure}[t]
\scriptsize
\fbox{
    \parbox{0.95\columnwidth}{         
        \textbf{\underline{HW Specification:}} Register report transmission due to \textbf{[T2]} and \textbf{[T1]}
        \begin{align*}
            \mathit{[T2]} &\mathit{\rightarrow t2_{pend}} \\
            \mathit{t2_{pend}} &\mathit{\land (read_{idx} = REPORT_{size}) \rightarrow \neg t2_{pend}} \\
            \mathit{[T1]} &\mathit{\rightarrow t1_{pend}} \\
            \mathit{t1_{pend}} &\mathit{\land \neg t2_{pend} \land (read_{idx} = REPORT_{size}) \rightarrow \neg t1_{pend}} \\
            \mathit{t1_{pend}} &\mathit{\lor t2_{pend} \rightarrow tx_{pend}}
        \end{align*}
        \textbf{\underline{HW Specification:}} Control transmission of \cflog
        \begin{equation*}
            \mathit{tx_{pend} \land \neg start_{\cflog} \land \vrf_{acc} \rightarrow start_{\cflog} \land (read_{idx} := \topslice)}
        \end{equation*}
        \begin{equation*}
            \mathit{start_{\cflog} \land (read_{idx} \neq REPORT_{size}) \rightarrow (read_{idx} := \texttt{incr}(read_{idx},1))}\\
        \end{equation*}
        \begin{equation*}
            \mathit{start_{\cflog} \land (read_{idx} = \bottomslice) \rightarrow finish_{\cflog} \land \neg start_{\cflog}}
        \end{equation*}
        
        \textbf{\underline{HW Specification}} Control transmission of remaining report data
        \begin{align*}
             & \mathit{finish_{\cflog} \land (PC = send_{addr}) \rightarrow start_{rem}} \land (read_{idx} := \cfsize) \\
             &\mathit{start_{rem} \land (read_{idx} \neq REPORT_{size}) \rightarrow (read_{idx} := read_{idx} + 1)} \\
             & \mathit{(read_{idx} = REPORT_{size}) \rightarrow \neg start_{rem} \land \neg finish}
        \end{align*}

        \textbf{\underline{HW Specification}} Assign next byte  as \textit{$data_{TX}$}.
        \begin{equation*} 
            \mathit{data_{TX}:= REPORT[read_{idx}]}
        \end{equation*}
    }
}
\vspace{-1.5em}
\caption{Contention Monitor transmission control}
\vspace{-2em} 
\label{fig:cont_mon_tx}
\end{figure}



        

\begin{figure}[t]
\scriptsize
\fbox{
    \parbox{0.95\columnwidth}{

        \textbf{\underline{HW Specification:}} Write received byte to \textit{RESPONSE}
        {
        \begin{align*}
            \mathit{trigger_{rx}} \rightarrow &\mathit{(RESPONSE[write_{idx}] := data_{RX})} \\
            &\land \mathit{(write_{idx} := write_{idx}+1)}
        \end{align*}
        }

        \textbf{\underline{HW Specification:}} Handle TCB Trigger \textbf{[T3]}
        {
        \begin{align*}
            \mathit{(write_{idx} = RESP_{Size})} &\rightarrow \mathit{resp_{pend} \land (write_{idx} := 0)} \\
            \mathit{\neg (write_{idx} = RESP_{Size})} &\rightarrow \mathit{\neg resp_{pend}}
        \end{align*}
        }
        
    } 
} 
\vspace{-1.5em}
\caption{Contention Monitor receive hardware specification}
\vspace{-2em}
\label{fig:cont_mon_rx}
\end{figure}

\subsection{Performance and Cost Evaluation}

\textit{\textbf{Memory Requirements.}}
\acron's  PMEM consists of the TCB software, which requires 5.3KB (4.5 KB for verified SHA-256-based HMAC). The remaining 830 bytes contain the CARAMEL workflow described in Sec.~\ref{sec:overview}.
%
%
%
\edit{\acron's DMEM size is determined by memory regions from Fig.~\ref{fig:arch}: \textit{METADATA}, \textit{REPORT}, and \textit{RESPONSE}.
For our prototype, we use 32-byte attestation challenges and authentication tokens (i.e., 32-byte HMAC). As a result, \metadata uses 38 bytes, and \textit{RESPONSE} uses 66 bytes.}
\edit{\acron's \textit{REPORT} requires 4 bytes for $\topslice$ and $\bottomslice$, 32 bytes for $\prv_{Auth}$, and the remaining size is based on the configuration of \cfsize, resulting in $\cfsize + 36$ bytes for \textit{REPORT}.
Therefore,
\acron's DMEM size is the sum of $\cfsize$ and all mentioned data sizes (140 bytes).
}

\textit{\textbf{Runtime Costs.}}
%
%
\acron aims to reduce the added runtime of \cfaud due to its busy-wait communication cycles.
To evaluate the improvement, we measure the runtime of audited executions of real-world publicly available sensor applications, namely:
an ultrasonic sensor application~\cite{ultsensor},
a temperature sensor application~\cite{tempsensor},
an automated medical syringe pump~\cite{syringepump},
and a motor control on a self-driving tank-tread rover~\cite{rover}.

We choose these applications because they were also used to evaluate the prior work~\cite{cflat,acfa,traces,oat}, allowing direct comparison. These applications are deployed atop \acron, \ACFA, and best-effort CFA. \edit{The latter provides no delivery guarantees and thus no ``busy wait'' overheads are associated with it}. Therefore, we refer to it as ``baseline''. We also report on two \acron settings for \cfsize: \cfsize of 2KB (\acron-2, i.e., two slices of 1KB each) and 4KB (\acron-4, i.e., two slices of 2KB each). \ACFA is set with a \cfsize of 2KB. The network latency for \prv$\leftrightarrow$\vrf round-trip time (RTT) is set to 100 $ms$, consistent with typical Internet RTTs~\cite{rtt}.

Fig.~\ref{fig:timing} presents the comparison of the runtime increase by the applications during an instance of audited execution. \edit{\ACFA increases runtime by $\approx$48.9-70.9$\%$, the highest compared to baseline, whereas \acron-2 only increases runtime by $\approx$24.8-41.3$\%$. With additional \cfsize, \acron-4 reduces this further to a $\approx$15.5-22.6$\%$ increase. 
The results indicate that \acron reduces the runtime compared to prior work \ACFA, and can reduce it further when more memory is available for \cflog.
}

\newcommand{\LUTcolor}{blue!50}
\newcommand{\FFcolor}{blue!10}
\newcommand{\baseline}{blue!80}
\newcommand{\firstconfig}{blue!40}
\newcommand{\secondconfig}{blue!20}

\begin{figure}[t]
    \centering
    \subfigure[\% Runtime increase vs. \cfatt \label{fig:timing}]
    {
        \resizebox{0.46\columnwidth}{!}{
            \begin{tikzpicture}
            \begin{axis} [
                xbar = .05cm,
            	bar width = 5pt,
                xmin=0,
                xmax=75,
            	xtick = {0,10,...,70},
            	ytick = data,
            	enlarge y limits = {abs = .8},
                height=.75\columnwidth,
                width=\columnwidth,
                ticklabel style = {font=\huge},
                yticklabels={\huge Ultra., \huge Temp., \huge Syringe., \huge Rover},
                legend style={at={(0,1.025)},anchor=south west},
                legend columns=3,
            ]
            

            \addplot[fill=\baseline] coordinates {
            (57.27,0) 
            (70.91,1) 
            (66.89,2) 
            (48.98,3) 
            };

            \addplot[fill=\secondconfig] coordinates {
            (24.83,0) 
            (41.3,1) 
            (36.75,2) 
            (40.95,3)  
            (0,3) 
            };

            \addplot[fill=white] coordinates {
            (16.38,0) 
            (22.57,1) 
            (20.99,2) 
            (15.50,3) 
            };
            
            \legend { \LARGE \ACFA, \LARGE \acron-2, \LARGE \acron-4};
            \end{axis}
            \end{tikzpicture}
            \vspace{-1em}
        }
    }
    \subfigure[\% Utilization increase vs. \ACFA\label{fig:utilization}]
    {
    \resizebox{0.46\columnwidth}{!}{
            \begin{tikzpicture}
            \begin{axis} [
                xbar = .05cm,
            	bar width = 5pt,
            	xmin = 0,
            	xmax = 155,
            	ytick = data,
            	enlarge y limits = {abs = .9},
                height=.75\columnwidth,
                width=\columnwidth,
                ticklabel style = {font=\huge},
                yticklabels={\huge Ultra., \huge Temp., \huge Syringe., \huge Rover},
                legend style={at={(0,1.025)},anchor=south west},
                legend columns=2
            ]

            \addplot[fill=\baseline] coordinates {
            (55.03,0) 
            (65,1) 
            (45.19,2) 
            (7.5,3)  
            };

            \addplot[fill=white] coordinates {
            (93.96,0) 
            (150,1) 
            (92.59,2) 
            (35.6,3) 
            };
            
            \legend { \huge \acron-2, \huge \acron-4};
            \end{axis}
            \end{tikzpicture}
            
        }
    }
    \vspace{-1.5em}
    \caption{Execution Runtime/Utilization vs. State-of-the-Art. 
    }
    \label{fig:costs}
    \vspace{-1.5em}
\end{figure}
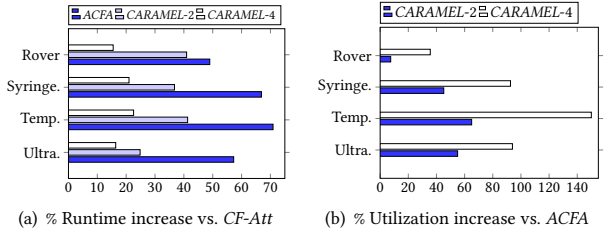

\edit{To evaluate the impact on CPU utilization, we compare the proportion of time spent executing the application with time spent executing \cfaud TCB code for both \acron configurations. Fig.~\ref{fig:utilization} shows the utilization compared to \ACFA. \acron-2 increases utilization across the selected applications by $\approx$7.4-65.0$\%$. As shown with \acron-4, a larger \cfsize gives an additional utilization increase for all applications. For \acron-4 the increased utilization ranges from $\approx$35.6-150$\%$.}


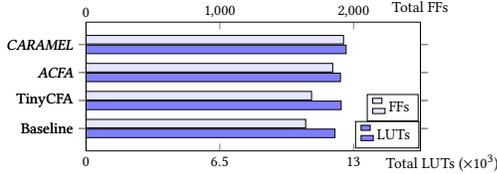
\begin{figure}[t]
    \centering
    \resizebox{0.8\columnwidth}{!}{
            \begin{tikzpicture}
            \begin{axis} [
                xbar = .05cm,
            	bar width = 5pt,
                bar shift = -0.10cm,
            	xmin = 0,
            	xmax = 13,
                xtick={0,6.5,13.0},
            	ytick = data,
            	enlarge x limits = {value = .25, upper},
            	enlarge y limits = {abs = .8},
                height=.5\columnwidth,
                width=\columnwidth,
                xlabel={Total LUTs ($\times 10^3$)},
                yticklabels={ Baseline, TinyCFA, \ACFA, \acron},
                axis x line*=bottom,
                legend style={at={(.805,0.02)},anchor=south west},
                label style={at={(axis description cs:1.25,-.2)},anchor=south east},
            ]
            \addplot[fill=\LUTcolor] coordinates {
            (12.097,0)
            (12.399,1)
            (12.363,2)
            (12.638,3)};
            \legend {LUTs};
            \end{axis}
            \begin{axis} [
                xbar = .05cm,
            	bar width = 5pt,
                bar shift = 0.10cm,
            	xmin = 0,
            	xmax = 2000,
                xtick={0,1000,2000},
            	ytick = data,
            	enlarge x limits = {value = .25, upper},
            	enlarge y limits = {abs = .8},
                height=.5\columnwidth,
                width=\columnwidth,
                xlabel={Total FFs },
                yticklabels={ Baseline,  TinyCFA,  \ACFA,  \acron},
                axis x line*=top,
                legend style={at={(.84,.23)},anchor=south west},
                label style={at={(axis description cs:1.1,1.2)},anchor=north east},
            ]
            \addplot[fill=\FFcolor] coordinates {
            (1642,0)
            (1686,1)
            (1844,2)
            (1925,3)
            };
            \legend {FFs};
            \end{axis}
            \end{tikzpicture}
        }
        \vspace{-1.5em}
        \caption{Hardware costs compared to related works}\label{fig:hw-detailed}
        \vspace{-1.5em}
\end{figure}


\emph{\textbf{Hardware Cost.}}
In line with prior work in this space~\cite{litehax,garota,lofat,pure,casu,Sancus17,acfa,apex,delegated,pfb,asap}, we measure the hardware cost based on the number of Look-up Tables (LUTs) and flip-flops/registers (FFs), as they indicate additional combinatorial logic cost and additional state, respectively. We present the cost of \acron and related works when deployed on a fully-fledged MCU chip, i.e., CPU core plus peripherals (timers, GPIO, UART, etc.).





The hardware cost results are shown in Fig.~\ref{fig:hw-detailed}. In total, \acron requires an additional 541 LUTs and 283 FFs, 4.5\% LUTs and 17.2\% FF over the unmodified openMSP430 core. 
Compared to ACFA, which provides \cfaud with less CPU utilization, \acron incurs an additional 2.2\% LUTs and 4.3\% FFs. 
This increase can be attributed most heavily to the CM-UART, which accounts for 4.1\% and 7.8\% of LUTs and FFs, in an \acron-equipped MCU.
Compared to a Tiny-CFA-equipped MCU, \acron adds 1.9\% LUTs and 14.2\% FFs. We note, however, that Tiny-CFA implements best effort delivery without guaranteeing communication between \vrf and \prv.

\subsection{Security Analysis}
Per \adv model in Sec. \ref{subsec:adv}, \adv can exploit software vulnerabilities on \prv, including modifying unprotected code or data (e.g., control flow hijack through the stack modification) and attempting to disrupt \acron protocol at any phase.

\edit{\acron control flow logging is hardware-based, hence, it cannot be altered by a software \adv. Thus, any control flow diversions caused by \adv exploits will be reflected in \cflog.
Instead, \adv may attempt to corrupt 
reports generated by \acron (e.g. differ from the actual content of \cflog). To achieve this, \adv would need to (1) alter TCB's execution, (2) replace \cfslice or the authentication token ($Prv_{Auth}$ from Fig.~\ref{fig:arch}), or (3) learn the secret key to forge reports.}
These options are prevented by \acron's inclusion of \ACFA as a sub-module, inheriting hardware protection for TCB code, \cflog, and keys~\cite{acfa}. \edit{\ACFA triggers a reset should these regions be illegally accessed. It also upholds immutability and controlled invocation of its TCB, ensuring it can only be entered or exited through dedicated points.
Additionally, data used by \acron (from Fig.~\ref{fig:arch}) are protected by its hardware from illegal modifications, making attacks that exploit new functionality infeasible.}


\edit{\adv may also attempt to forge/modify \vrf response to avoid remediation. 
Since all incoming messages are authenticated within TCB, this is infeasible.
Alternatively, \adv may block \vrf's communication from reaching \prv. In this case, \acron will eventually trigger its TCB through \textbf{[T4]}, entering a secure software state while trying to retransmit \cfslices until an authenticated response from \vrf is received. This ensures that reports are not lost as long as there is eventual communication. Lastly, \adv attempts to disable/misconfigure the CM-UART (e.g., tampering the transmission rate) are also prevented by \acron hardware write protection.}

\section{Related Work}




\edit{
\textit{\textbf{Runtime Attestation.}}
%
%
\cfatt, combining TEE hardware support (from ARM TrustZone) with code instrumentation, was first introduced in C-FLAT~\cite{cflat}.
 In C-FLAT, branch instructions are instrumented with a \textit{trampolines} to the TrustZone-protected ``Secure World'' to build a hash-chain of branch destinations uniquely representing the path followed during execution. 
Several techniques~\cite{scarr,recfa,oat,ari,traces,blast,enola} build upon C-FLAT combining some form of instrumentation and TEE support. They trade-off instrumentation for added hardware cost. 
}

\edit{\textit{\textbf{Runtime Auditing.}} \ACFA~\cite{acfa} originally proposed \cfaud by augmenting \cfatt with "active root of trust"~\cite{garota} to eventually deliver runtime evidence to \vrf (assuming the network is not blocked indefinitely). \ACFA uses custom hardware to build \cflog and actively trigger execution of its trusted software. The trusted software generates and transmits the report. TRACES~\cite{traces} demonstrated that \cfatt is achievable without requiring custom hardware. It uses instrumentation and TEE support to instantiate the same guarantees of \ACFA. After sending reports, \ACFA and TRACES require \prv to wait in a secure state before continuing execution, decreasing \prv utilization. \acron aims to address this problem.}

\edit{\textit{\textbf{Reliable Data Transfer.}}
Network-level reliable data transfer~\cite{stann2003rmst,patra2003dtnlite,akan2005event, wan2002psfq,deb2003reinform} is well-studied in the context where both endpoints are honest. Within our context, \acron creates a root of trust with a self-contained communication capability. Different from prior work in this space, this ensures reliable delivery despite software compromise of the endpoint where this RoT resides.}

\section{Conclusion}

We present \acron, a \cfaud architecture designed to boost \prv utilization by mitigating busy-wait cycles as part of its reliable evidence delivery protocol. At its core, \acron introduces a novel RoT embedded with a self-contained and reliable communication interface. We developed an open-source prototype of \acron~\cite{caramelrepo}, and our evaluation shows that it significantly improves \prv utilization compared to the state-of-the-art, while incurring modest hardware overhead.





\bibliographystyle{ACM-Reference-Format}
\bibliography{references} 

@article{syringepump,
title = {An 8051 Microcontroller based Syringe Pump Control System for Surface Micromachining},
note = {ICAMME},
author = {M.S.V. Appaji and others},
}

@article{enola,
  title={{ENOLA}: Efficient Control-Flow Attestation for Embedded Systems},
  author={Armanuzzaman, Md and Kirda, Engin and Zhao, Ziming},
  journal={arXiv preprint arXiv:2501.11207},
  year={2025}
}

@inproceedings{speccfa,
  author       = {Adam Caulfield and
                  Liam Tyler and
                  Ivan De Oliveira Nunes},
  title        = {SpecCFA: Enhancing Control Flow Attestation/Auditing via Application-Aware
                  Sub-Path Speculation},
  booktitle    = {Annual Computer Security Applications Conference, {ACSAC} 2024},
  pages        = {563--578},
  publisher    = {{IEEE}},
  year         = {2024}
}

@inproceedings{iscflat,
  title={ISC-FLAT: On the Conflict Between Control Flow Attestation and Real-Time Operations},
  author={Neto, Antonio Joia and Nunes, Ivan De Oliveira},
  booktitle={2023 IEEE 29th Real-Time and Embedded Technology and Applications Symposium (RTAS)},
  pages={133--146},
  year={2023},
  organization={IEEE}
}

@inproceedings{jop,
  title={Jump-oriented programming: a new class of code-reuse attack},
  author={Bletsch, Tyler and others},
  booktitle={ACM CCS},
  pages={30--40},
  year={2011}
}

@inproceedings{recfa,
  title={{ReCFA}: resilient control-flow attestation},
  author={Zhang, Yumei and Liu, Xinzhi and Sun, Cong and Zeng, Dongrui and Tan, Gang and Kan, Xiao and Ma, Siqi},
  booktitle={Annual Computer Security Applications Conference},
  pages={311--322},
  year={2021}
}

@inproceedings {acfa,
author = {Adam Caulfield and Norrathep Rattanavipanon and Ivan De Oliveira Nunes},
title = {{ACFA}: Secure Runtime Auditing \& Guaranteed Device Healing via Active Control Flow Attestation},
booktitle = {32nd USENIX Security Symposium (USENIX Security 23)},
year = {2023},
pages = {5827--5844}
}

@misc{opensyringe,
  title =  {{OpenSyringePump}},
  month = apr,
  author = {Theo Walker},
  url = {https://github.com/manimino/OpenSyringePump},
  year =   {2022},
}

@misc{tempsensor,
  title =  {{Temperature Sensor}},
  author = {Seeed-Studio},
  url = {https://github.com/Seeed-Studio/LaunchPad\_Kit/tree/master/Grove\_Modules/temp\_humi\_sensor},
  month = jun,
  year =   2015,
}

@misc{ultsensor,
  title =  {{Ultrasonic Ranger}},
  author = {Seeed-Studio},
  url = {https://github.com/Seeed-Studio/LaunchPad\_Kit/tree/master/Grove\_Modules/ultrasonic\_ranger},
  month = jun,
  year =   2015,
}

@inproceedings{delegated,
  title={Delegated attestation: scalable remote attestation of commodity cps by blending proofs of execution with software attestation},
  author={Ammar, Mahmoud and Crispo, Bruno and De Oliveira Nunes, Ivan and Tsudik, Gene},
  booktitle={Proceedings of the 14th ACM Conference on Security and Privacy in Wireless and Mobile Networks},
  pages={37--47},
  year={2021}
}

@article{obermaier2018past,
  title={The past, present, and future of physical security enclosures: from battery-backed monitoring to PUF-based inherent security and beyond},
  author={Obermaier, Johannes and Immler, Vincent},
  journal={Journal of Hardware and Systems Security},
  volume={2},
  number={4},
  pages={289--296},
  year={2018},
  publisher={Springer}
}

@inproceedings{scarr,
  title={{ScaRR}: Scalable Runtime Remote Attestation for Complex Systems},
  author={Toffalini, Flavio and Losiouk, Eleonora and Biondo, Andrea and Zhou, Jianying and Conti, Mauro},
  booktitle={22nd International Symposium on Research in Attacks, Intrusions and Defenses (RAID 2019)},
  pages={121--134},
  year={2019}
}

@article{aliasing,
  title={The undecidability of aliasing},
  author={Ramalingam, Ganesan},
  journal={ACM Transactions on Programming Languages and Systems (TOPLAS)},
  volume={16},
  number={5},
  pages={1467--1471},
  year={1994},
  publisher={ACM New York, NY, USA}
}

@inproceedings{garota,
  title={{GAROTA}: Generalized Active Root-Of-Trust Architecture (for Tiny Embedded Devices)},
  author={Aliaj, Esmerald and Nunes, Ivan De Oliveira and Tsudik, Gene},
  booktitle={31st USENIX Security Symposium (USENIX Security 22)},
  pages={2243--2260},
  year={2022}
}

@inproceedings{asap,
  title={{ASAP}: reconciling asynchronous real-time operations and proofs of execution in simple embedded systems},
  author={Caulfield, Adam and Rattanavipanon, Norrathep and De Oliveira Nunes, Ivan},
  booktitle={Proceedings of the 59th ACM/IEEE Design Automation Conference},
  pages={721--726},
  year={2022}
}

@inproceedings{tinycfa,
  title={{Tiny-CFA}: Minimalistic control-flow attestation using verified proofs of execution},
  author={Nunes, Ivan De Oliveira and Jakkamsetti, Sashidhar and Tsudik, Gene},
  booktitle={2021 Design, Automation \& Test in Europe Conference \& Exhibition (DATE)},
  pages={641--646},
  year={2021},
  organization={IEEE}
}

@article{sok_cfa_cfi,
  title={SoK: Runtime Integrity},
  author={Ammar, Mahmoud and Caulfield, Adam and Nunes, Ivan De Oliveira},
  journal={arXiv preprint arXiv:2408.10200}  ,
  year={2024}
}

@article{cfi-survey-1,
  title={Control-flow integrity: Precision, security, and performance},
  author={Burow, Nathan and Carr, Scott A and Nash, Joseph and Larsen, Per and Franz, Michael and Brunthaler, Stefan and Payer, Mathias},
  journal={ACM Computing Surveys (CSUR)},
  volume={50},
  number={1},
  pages={1--33},
  year={2017},
  publisher={ACM New York, NY, USA}
}

@inproceedings{casu,
  title={CASU: Compromise Avoidance via Secure Update for Low-end Embedded Systems},
  author={De Oliveira Nunes, Ivan and Jakkamsetti, Sashidhar and Kim, Youngil and Tsudik, Gene},
  booktitle={Proceedings of the 41st IEEE/ACM International Conference on Computer-Aided Design},
  pages={1--9},
  year={2022}
}

@inproceedings{cflat,
  title={C-FLAT: control-flow attestation for embedded systems software},
  author={Abera, Tigist and others},
  booktitle={SIGSAC},
  pages={743--754},
  year={2016}
}

@inproceedings{vrased,
  title={{VRASED}: A Verified {Hardware/Software Co-Design} for Remote Attestation},
  author={Nunes, Ivan De Oliveira and Eldefrawy, Karim and Rattanavipanon, Norrathep and Steiner, Michael and Tsudik, Gene},
  booktitle={28th USENIX Security Symposium (USENIX Security 19)},
  pages={1429--1446},
  year={2019}
}

@inproceedings{apex,
  title={{APEX}: A verified architecture for proofs of execution on remote devices under full software compromise},
  author={Nunes, Ivan De Oliveira and Eldefrawy, Karim and Rattanavipanon, Norrathep and Tsudik, Gene},
  booktitle={29th USENIX Security Symposium (USENIX Security 20)},
  pages={771--788},
  year={2020}
}

@article{rop,
  title={Return-oriented programming: Systems, languages, and applications},
  author={Roemer, Ryan and others},
  journal={TISSEC},
  volume={15},
  number={1},
  pages={1--34},
  year={2012},
  publisher={ACM New York, NY, USA}
}

@inproceedings{oat,
  title={OAT: Attesting operation integrity of embedded devices},
  author={Sun, Zhichuang and Feng, Bo and Lu, Long and Jha, Somesh},
  booktitle={2020 IEEE Symposium on Security and Privacy (SP)},
  pages={1433--1449},
  year={2020},
  organization={IEEE}
}

@article{Sancus17,
  title={Sancus 2.0: A low-cost security architecture for iot devices},
  author={Noorman, Job and Bulck, Jo Van and M{\"u}hlberg, Jan Tobias and Piessens, Frank and Maene, Pieter and Preneel, Bart and Verbauwhede, Ingrid and G{\"o}tzfried, Johannes and M{\"u}ller, Tilo and Freiling, Felix},
  journal={ACM Transactions on Privacy and Security (TOPS)},
  volume={20},
  number={3},
  pages={1--33},
  year={2017},
  publisher={ACM New York, NY, USA}
}

@misc{openmsp430,
  title={open{MSP430}},
  author={Girard, Olivier},
  year={2009},
  howpublished = {\url{https://opencores.org/projects/openmsp430}},
  publisher={OpenCores}
}

@inproceedings{lofat,
  title={Lo-fat: Low-overhead control flow attestation in hardware},
  author={Dessouky, Ghada and Zeitouni, Shaza and Nyman, Thomas and Paverd, Andrew and Davi, Lucas and Koeberl, Patrick and Asokan, N and Sadeghi, Ahmad-Reza},
  booktitle={Proceedings of the 54th Annual Design Automation Conference 2017},
  pages={1--6},
  year={2017}
}

@inproceedings{atrium,
  title={Atrium: Runtime attestation resilient under memory attacks},
  author={Zeitouni, Shaza and Dessouky, Ghada and Arias, Orlando and Sullivan, Dean and Ibrahim, Ahmad and Jin, Yier and Sadeghi, Ahmad-Reza},
  booktitle={2017 IEEE/ACM International Conference on Computer-Aided Design (ICCAD)},
  pages={384--391},
  year={2017},
  organization={IEEE}
}

@inproceedings{hacl,
  title={HACL*: A verified modern cryptographic library},
  author={Zinzindohou{\'e}, Jean-Karim and Bhargavan, Karthikeyan and Protzenko, Jonathan and Beurdouche, Benjamin},
  booktitle={Proceedings of the 2017 ACM SIGSAC Conference on Computer and Communications Security},
  pages={1789--1806},
  year={2017},
  organization={ACM}
}

@inproceedings{ari,
  title={{ARI}: Attestation of Real-time Mission Execution Integrity},
  author={Wang, Jinwen and Wang, Yujie and Li, Ao and Xiao, Yang and Zhang, Ruide and Lou, Wenjing and Hou, Y Thomas and Zhang, Ning},
  booktitle={32nd USENIX Security Symposium (USENIX Security 23)},
  pages={2761--2778},
  year={2023}
}

@inproceedings{ma2023ret2ns,
  title={Return-to-Non-Secure Vulnerabilities on ARM Cortex-M TrustZone: Attack and Defense},
  author={Ma, Zheyuan and Tan, Xi and Ziarek, Lukasz and Zhang, Ning and Hu, Hongxin and Zhao, Ziming},
  booktitle={2023 60th ACM/IEEE Design Automation Conference (DAC)},
  pages={1--6},
  year={2023},
  organization={IEEE}
}

@inproceedings{pure,
  title={Pure: Using verified remote attestation to obtain proofs of update, reset and erasure in low-end embedded systems},
  author={De Oliveira Nunes, Ivan and Eldefrawy, Karim and Rattanavipanon, Norrathep and Tsudik, Gene},
  booktitle={2019 IEEE/ACM International Conference on Computer-Aided Design (ICCAD)},
  pages={1--8},
  year={2019},
  organization={IEEE}
}

@inproceedings{traces,
  author       = {Adam Caulfield and
                  Antonio Joia Neto and
                  Norrathep Rattanavipanon and
                  Ivan De Oliveira Nunes},
  title        = {{TRACES:} TEE-based Runtime Auditing for Commodity Embedded Systems},
  booktitle    = {Annual Computer Security Applications Conference, {ACSAC} 2024},
  pages        = {257--270},
  publisher    = {{IEEE}},
  year         = {2024}
}

@inproceedings{blast,
  title={Whole-Program Control-Flow Path Attestation},
  author={Yadav, Nikita and Ganapathy, Vinod},
  booktitle={Proceedings of the 2023 ACM SIGSAC Conference on Computer and Communications Security},
  pages={2680--2694},
  year={2023}
}

@inproceedings{litehax,
  title={Litehax: lightweight hardware-assisted attestation of program execution},
  author={Dessouky, Ghada and Abera, Tigist and Ibrahim, Ahmad and Sadeghi, Ahmad-Reza},
  booktitle={2018 IEEE/ACM International Conference on Computer-Aided Design (ICCAD)},
  pages={1--8},
  year={2018},
  organization={IEEE}
}

@inproceedings{pfb,
  title={Privacy-from-birth: Protecting sensed data from malicious sensors with VERSA},
  author={Nunes, Ivan De Oliveira and others},
  booktitle={2022 IEEE Symposium on Security and Privacy (SP)},
  pages={2413--2429},
  year={2022},
  organization={IEEE}
}

@inproceedings{rtt,
  title={Measurements on delay and hop-count of the internet},
  author={Fei, Aiguo and Pei, Guangyu and Liu, Roy and Zhang, Lixia},
  booktitle={IEEE GLOBECOM’98-Internet Mini-Conference},
  year={1998}
}

@misc{caramelrepo,
author = {Lengert, Alexandra and Caulfield, Adam and De Oliveira Nunes, Ivan},
month = March,
title =  {{\acron Github Repository}},
url = {https://github.com/SPINS-RG/CARAMEL},
year = {2026}
}

@misc{rover,
  title =  {{Rover}},
  author = {RiS3-Lab},
  url = {https://github.com/RiS3-Lab/OAT-Project/blob/master/oat-evaluation/roverpi-orig/rovertcp.c},
  month = nov,
  year =   2020,
}

@inproceedings{francillon2008code,
  title={Code injection attacks on harvard-architecture devices},
  author={Francillon, Aur{\'e}lien and others},
  booktitle={Proceedings of the 15th ACM conference on Computer and communications security},
  pages={15--26},
  year={2008}
}

@inproceedings{soltan2018blackiot,
  title={{BlackIoT}: {IoT} botnet of high wattage devices can disrupt the power grid},
  author={Soltan, Saleh and others},
  booktitle={27th USENIX security symposium (USENIX security 18)},
  pages={15--32},
  year={2018}
}

@inproceedings{alrawi2019sok,
  title={Sok: Security evaluation of home-based iot deployments},
  author={Alrawi, Omar and others},
  booktitle={2019 IEEE symposium on security and privacy (sp)},
  pages={1362--1380},
  year={2019},
  organization={IEEE}
}

@inproceedings{shekari2021mamiot,
  title={{MaMIoT}: Manipulation of energy market leveraging high wattage IoT botnets},
  author={Shekari, Tohid and others},
  booktitle={Proceedings of the 2021 ACM SIGSAC Conference on Computer and Communications Security},
  pages={1338--1356},
  year={2021}
}

@article{nunes2024toward,
  title={Toward Remotely Verifiable Software Integrity in Resource-Constrained IoT Devices},
  author={Nunes, Ivan De Oliveira and others},
  journal={IEEE Communications Magazine},
  volume={62},
  number={7},
  pages={58--64},
  year={2024},
  publisher={IEEE}
}

@inproceedings{stann2003rmst,
  title={RMST: Reliable data transport in sensor networks},
  author={Stann, Fred and others},
  booktitle={Proceedings of the First IEEE International Workshop on Sensor Network Protocols and Applications, 2003.},
  pages={102--112},
  year={2003},
  organization={IEEE}
}

@article{patra2003dtnlite,
  title={Dtnlite: A reliable data transfer architecture for sensor networks},
  author={Patra, Rabin and others},
  journal={CS294-1: Deeply Embedded Networks (Fall 2003)},
  year={2003}
}

@article{akan2005event,
  title={Event-to-sink reliable transport in wireless sensor networks},
  author={Akan, Ozg{\"u}r B and others},
  journal={IEEE/ACM transactions on networking},
  volume={13},
  number={5},
  pages={1003--1016},
  year={2005},
  publisher={IEEE}
}

@inproceedings{wan2002psfq,
  title={PSFQ: a reliable transport protocol for wireless sensor networks},
  author={Wan, Chieh-Yih and others},
  booktitle={Proceedings of the 1st ACM international workshop on Wireless sensor networks and applications},
  pages={1--11},
  year={2002}
}

@inproceedings{deb2003reinform,
  title={ReInForM: Reliable information forwarding using multiple paths in sensor networks},
  author={Deb, Budhaditya and others},
  booktitle={28th Annual IEEE International Conference on Local Computer Networks, 2003. LCN'03. Proceedings.},
  pages={406--415},
  year={2003},
  organization={IEEE}
}

@misc{tiMSP430,
	title = {MSP430-Mikrocontroller},
    author = {Texas Instruments Inc.},
    year ={2025}, 
	url = {https://www.ti.com/de-de/product-category/microcontrollers-processors/msp430-mcus/overview.html},
	urldate = {2025-11-18}
}

@misc{atmega,
	title = {ATmega Microcontrollers (MCUs)},
    author = {Microchip Technology Inc.},
    year ={2025}, 
	url = {https://www.microchip.com/en-us/about/corporate-overview/acquisitions/atmel/atmega},
	urldate = {2025-11-18}
}

\end{document}